\documentclass[pra,aps,twocolumn,10pt,groupedaddress,superscriptaddress,floatfix]{revtex4-2}
\usepackage{amsmath,amsfonts,amssymb,graphics,graphicx,epsfig,color,times}
\usepackage[utf8]{inputenc} 
\usepackage{color}
\usepackage{bbm, dsfont} 
\usepackage{subfigure}
\usepackage{hyperref}
\usepackage{mathrsfs}
\usepackage{verbatim}
\usepackage{braket}
\begin{document}

\title{Almost indistinguishable single photons via multiplexing cascaded biphotons with cavity modulation and phase compensation}

\author{Y.-E. Wong}
\email{b06504045@ntu.edu.tw}
\affiliation{Department of Mechanical Engineering, National Taiwan University, Taipei 106319, Taiwan}

\author{T. H. Chang}
\affiliation{Institute of Atomic and Molecular Sciences, Academia Sinica, Taipei 10617, Taiwan}

\author{H. H. Jen}
\email{sappyjen@gmail.com}
\affiliation{Institute of Atomic and Molecular Sciences, Academia Sinica, Taipei 10617, Taiwan}
\affiliation{Physics Division, National Center for Theoretical Sciences, Taipei 10617, Taiwan}

\date{\today}
\renewcommand{\k}{\mathbf{k}}
\renewcommand{\r}{\mathbf{r}}
\newcommand{\parallelsum}{\mathbin{\!/\mkern-5mu/\!}}
\def\p{\mathbf{p}}
\def\R{\mathbf{R}}
\def\bea{\begin{eqnarray}}
\def\eea{\end{eqnarray}}
\def\bee{\begin{equation}}
\def\eee{\end{equation}}
\begin{abstract}
The cascade-emitted biphotons generated from the alkali metal atomic ensembles are an excellent entanglement resource which enables long-distance quantum communication. The communication of quantum information between distant locations can be realized by utilizing the low-loss telecom bandwidth in the upper transition of the cascaded photons in a fiber-based quantum network. Meanwhile, the infrared photon from the lower transition of this highly directional and frequency-correlated biphoton can be created under the four-wave mixing process and can be stored locally as a collective spin wave. Here we theoretically investigate the frequency entanglement of this biphoton and propose two approaches to remove their mutual correlations in frequency spaces. The first approach applies an optical cavity which modulates the biphoton spectrum into a more symmetric and narrow spectral function by multiplexing multiple atomic ensembles with phase compensation. The purity of single photon reaches up to $0.999$ and the entanglement entropy $S$ of the biphoton reduces to $0.006$, which is $200$ times smaller than the one without multiplexing. The other approach employs a symmetric pumping of the laser fields in two atomic ensembles, which leads to a moderate reduction of $S\sim 0.3$ when non-discrimination detection devices are used for both photons. An extremely low frequency entanglement implies an almost indistinguishable single photon source, which offers a potential resource for photonic quantum simulation and computation.
\end{abstract}
\maketitle
\section{Introduction}

Quantum communication has been actively developed and successfully implemented in many diverse physical systems, including neutral atoms, trapped ions, superconducting qubits, quantum dots, and diamond color centers. It has became an essential building block in constructing the quantum network \cite{Cirac1997, Kimble2008}, where the interconnection nodes formed by these various platforms can be integrated \cite{Awschalom2021}. A recent advancement in long-distance quantum communication \cite{Duan2001} has been pushed to Earth-orbit satellite distance \cite{Yin2017, Liao2017, Liao2018, YAChen2021}, which promises a global quantum communication by integrating metropolitan-area networks with the ground-satellite links. In quantum communication, a quantum repeater protocol \cite{Briegel1998, Dur1999, Sangouard2011} serves as an elementary functionality to genuinely relay entanglement to distant locations before it attenuates by dissipations in a fiber link. In this fiber-based communication scheme, an efficient quantum interface between light and atoms \cite{Hammerer2010, Chen2013, Yang2016, Hsiao2018} is therefore demanded, and a telecom bandwidth from the atomic cascaded emissions \cite{Chaneliere2006, Radnaev2010, Jen2010} can further reduce the losses in fiber transmission.       

In addition to the improvements of utilizing high-efficiency quantum interface and suitable bandwidth in communication channels, a multiplexing scheme of quantum memories in space \cite{Collins2007, Lan2009}, time \cite{Simon2007} or frequencies \cite{Bernhard2013, Lukens2014, Jen2016a, Jen2016b, Lukens2017, Kues2017, Chang2020, Shi2020} can further allow a higher communication capacity of entanglement in respective continuous spaces, leading to a multimode quantum information processing \cite{Afzelius2009, Dai2012, Zheng2015}. This is in contrast to the system using finite and discrete degrees of freedom, where paths, polarization, or orbital angular momenta can be employed for entanglement generation \cite{Wang2018_Pan, Chen_Liu2021}. On the other hand, a low frequency-correlated biphoton source can host almost pure heralded single photons, which can be applied to implement linear optical quantum networks \cite{Dusanowski2019}, quantum computation with linear optics \cite{Knill2001}, and photon–photon quantum logic gates \cite{Li2021}.  

\begin{figure}[b]
\centering
\includegraphics[width=8.5cm,height=4.5cm]{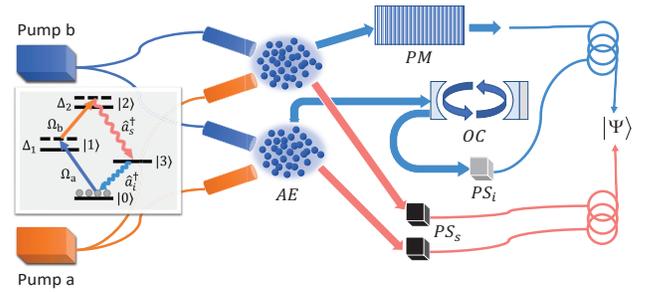}
\caption{A schematic setup of spectral manipulation in biphoton state via multiplexing two atomic ensembles (AEs) with cavity modulation and phase compensation. Two pumping fields $\Omega_a$ and $\Omega_b$ generate signal ($\hat a_s^\dag$) and idler photons ($\hat a_i^\dag$) sequentially under the four-wave mixing condition. The idler photons go through either an optical cavity (OC) or phase modulator (PM) to form an effective spectral function with a reduced entanglement entropy $S$. Phase shifter (PS) is applied for signal or idler photons for exploration and manipulation of $S$, and $|\Psi\rangle$ represents an effective biphoton state function. }\label{fig1}
\end{figure}

Here we consider cascaded atomic configurations as shown in Fig. \ref{fig1}, where frequency-correlated signal and idler photons can be generated under four-wave mixing condition. We propose two approaches to remove the frequency correlations in the cascaded biphoton. The first approach utilizes an optical cavity to modulate the biphoton spectrum by multiplexing multiple atomic ensembles with phase compensation. A reduction of the frequency entanglement entropy emerges owing to a more symmetric and narrow frequency distribution. The other approach uses a symmetric pumping of the laser fields in two atomic ensembles, which leads to a moderate reduction of entanglement entropy when non-discrimination detection devices are considered for both photons. An extremely low entanglement entropy in frequency spaces suggests an almost identical single photon source with near-unity purity, which provides a tunable resource for quantum simulation and computation with photons.

The paper is organized as follows. In Sec. II, we introduce the biphoton spectral function and its entanglement entropy in frequency spaces determined by Schmidt decomposition. We then investigate the modification of entanglement entropy and associated purity of single photons in a multiplexing scheme with cavity modulation and phase compensation in Sec. III. In Sec. IV, we further demonstrate a method of symmetrizing the spectral function and its associated entanglement property, and we conclude in Sec. V. 

\section{Spectral function and frequency entanglement}

The cascaded and frequency-correlated biphoton state in Fig. \ref{fig1} can be generated via two pump fields under the four-wave mixing condition. Under the weak laser field excitation, we obtain the effective biphoton state
\bea
|\Psi\rangle\approx f(\omega_s,\omega_i)\hat a_{\k_s}^\dag\hat a_{\k_i}^\dag|0\rangle, 
\eea
where $\k_{s}$ and $\k_i$ indicate the wave vectors of the signal and idler photons, respectively, and its spectral function can be written as  
\begin{equation}
\label{singleF}
f(\omega_s,\omega_i)=\frac{e^{-(\Delta\omega_s+\Delta\omega_i)^2\tau^2/8}}{\Gamma_3^N/2-i\Delta\omega_i},
\end{equation}
which we obtain with detail in Appendix A. The $\tau$ represents a pulse duration from a laser excitation in a Gaussian wave form, and the $\Gamma_3^N$ is superradiant decay constant for the idler transition \cite{Chaneliere2006}. The frequency distributions for signal and idler photons are $\Delta \omega_s \equiv \omega_s - (\omega_2 - \omega_3 +\Delta_2)$ and $\Delta \omega_i \equiv \omega_i - \omega_3$, respectively. The spectral function of Eq. (\ref{singleF}) demonstrates a Lorentzian idler photon distribution with an enhanced decay rate and a joint Gaussian distribution for both signal and idler photons. It is this joint distribution that contributes to the continuous entanglement entropy in frequency spaces, which increases when the anti-correlation distribution along $\Delta\omega_s+\Delta\omega_i=0$ dominates. 

Next we express the effective biphoton state in an integral form in frequency spaces, 
\bea
|\Psi'\rangle=\mathcal{N}\int\int f(\omega_s,\omega_i)\hat a_{\k_s}^\dag(\omega_s)\hat a_{\k_i}^\dag(\omega_i)|0\rangle d\omega_sd\omega_i,\label{psiprime}
\eea
where $\mathcal{N}$ represents the normalization constant. This integral form provides a recipe to calculate the entanglement entropy in continuous frequency spaces \cite{Branning1999, Law2000, Parker2000, Jen2012-2}. The continuous entanglement entropy can then be obtained in the Schmidt bases, where the state vectors can be rewritten as
\bea
|\Psi'\rangle=&&\sum_{n}\sqrt{\lambda_{n}}\hat{b}_{n}^{\dag}\hat{c}_{n}^{\dag}|0\rangle,\\
\hat{b}_{n}^{\dag}\equiv&&\int\psi_{n}(\omega_{s})\hat{a}_{\k_{s}}^{\dag}(\omega_{s})d\omega_{s},\\
\hat{c}_{n}^{\dag}\equiv&&\int\phi_{n}(\omega_{i})\hat{a}_{\k_{i}}^{\dag}(\omega_{i})d\omega_{i}.
\eea
In the Schmidt bases, $\lambda_n$ represents the state probability for the $n$th signal and idler eigenmodes $\psi_{n}$ and $\phi_{n}$, respectively. These eigenmodes and eigenvalues can be calculated from the kernels $K_{1,2}$ which satisfy 
\begin{eqnarray}
&&\int K_{1}(\omega,\omega^{\prime})\psi_{n}(\omega^{\prime})d\omega^{\prime}  =\lambda_{n}\psi_{n}(\omega),\\
&&\int K_{2}(\omega,\omega^{\prime})\phi_{n}(\omega^{\prime})d\omega^{\prime}  =\lambda_{n}\phi_{n}(\omega),
\end{eqnarray}
where $K_{1,2}$ can be constructed from the one-photon spectral correlations \cite{Law2000, Parker2000},
\begin{eqnarray}
&&K_{1}(\omega,\omega^{\prime}) \equiv\int f(\omega,\omega_{1})f^{\ast}(\omega^{\prime},\omega_{1})d\omega_{1},\\
&&K_{2}(\omega,\omega^{\prime}) \equiv\int f(\omega_{2},\omega)f^{\ast}(\omega_{2},\omega^{\prime})d\omega_{2}. 
\end{eqnarray}
We note that the orthogonality relations of these eigenmodes should be satisfied as $\int\psi_{i}(\omega)$$\psi_{j}^*(\omega)d\omega$ $=$ $\delta_{ij}$ and $\int\phi_{i}(\omega)$$\phi_{j}^*(\omega)d\omega$ $=$ $\delta_{ij}$, and $\sum_{n}\lambda_{n}$ should be normalized to one, showing the completeness of the Schmidt bases. 

From the above Schmidt decomposition, we can continue to calculate the entanglement entropy $S$ in frequency spaces as
\begin{eqnarray}
S=-\sum_{n=1}^{\infty}\lambda_{n}\textrm{log}_2\lambda_{n}.\label{entropy}
\end{eqnarray}
When only one $\lambda_n$ is finite, that is $\lambda_{1}=1$, $S$ is vanishing, which attributes to a separable or non-entangled state. Other than this, a finite $S>0$ indicates an entangled biphoton source, which results from more than one nonzero Schmidt numbers $\lambda_n$ in $|\Psi'\rangle$. In the following sections, we use $S$ as a measure for continuous entanglement entropy, and locating its minimum can be a guide to identify the parameter regimes to host an almost indistinguishable single photons from this biphoton source. The associated degree of single photon purity can also be calculated as Tr$(\rho_{i(s)}^2)$ where $\rho_{i(s)}$ $\equiv$ Tr$_{s(i)}(\rho)$, $\rho$ $\equiv$ $|\Psi'\rangle\langle\Psi'|$, and the subscripts $i(s)$ represents the idler (signal) frequency degrees of freedom in Eq. (\ref{psiprime}). 

\section{Multiplexing scheme with cavity modulation and phase compensation}

The capacity of continuous entanglement entropy in frequency space can be increased by multiplexing multiple atomic ensembles \cite{Jen2016a} along with phase modulations \cite{Law2000, Jen2016b}. This multiplexing scheme has been demonstrated with superiority of spectral manipulations in multimode quantum memory \cite{Yang2018} and enhanced performance in quantum dense coding \cite{Chen2021}. In addition to the multiplexing scheme that can shape the spectral property of the biphoton state, an optical cavity can further spectrally modulate the spectral distribution and effectively compress the linewidth of the photons \cite{Seidler2020, Wong2021}. Therefore, as shown in Fig. \ref{fig1}, we propose to spectrally manipulate the spectral function of the cascaded biphoton state by multiplexing multiple atomic ensembles with the optical elements of phase modulator or optical cavity. Specifically, an optical cavity can be formed by a two-mirror cavity with one side of unit reflectivity, which gives rise to a lossless transfer function \cite{Agarwal1994, Srivathsan2014, Liu2014},  
\bea
\label{cavityEq}        
C_t(\Delta\omega,\Gamma_c) = -\frac{\Gamma_c + i2\Delta\omega}{\Gamma_c - i2\Delta\omega},
\eea
where $\Gamma_c$ characterizes the cavity linewidth. The above transformation essentially imprints a continuous phase from $0$ to $\pi$ as $\Delta\omega$ changes from a large spread toward a resonant case of $\Delta\omega\rightarrow 0$ as $\Delta\omega$ approaches the range around $\pm \Gamma_c/2$. In the limit of $\Gamma_c\rightarrow\infty$, Eq. (\ref{cavityEq}) in general imprints an overall constant phase of $\pi$ without modulating the spectrum. 

\begin{figure}[b]
\centering
\includegraphics[width=8.5cm,height=4.5cm]{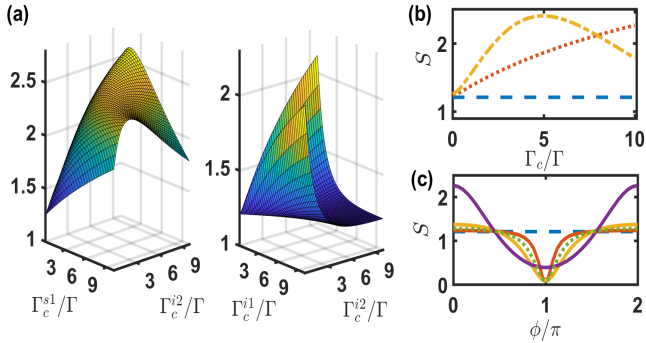}
\caption{The effect of optical cavity linewidth and relative phase in entanglement entropy $S$. (a) In the left panel, the change of $S$ for the first signal and the second idler photons going through the cavities with $\Gamma_c^{s1}$ and $\Gamma_c^{i2}$, respectively. The case for $S$ when both idler photons going through respective cavities with $\Gamma_c^{i1}$ and $\Gamma_c^{i2}$ in the right panel. (b) Two cuts in the left panel of (a) for $(\Gamma_c^{s1}, \Gamma_c^{i2})=(0,\Gamma_c)$ (red-dotted line) and $\Gamma_c^{s1}=\Gamma_c^{i2}=\Gamma_c$ (yellow dash-dotted line), comparing the cavity-free case (blue dashed line). (c) The dependence of $S$ in the phase $\phi$ of the first idler photon with only the second idler photon through a cavity of $\Gamma_c^{i2}/\Gamma$ $=$ $0.1$, $1$, $10$ (solid lines), where $S$ increases as $\Gamma_c^{i2}$ increases at $\phi=0$, or with only the second signal photon of $\Gamma_c^{s2}/\Gamma$ $=$ $1$ (dotted line) as a comparison. The spectral range for calculating $S$ is taken as $\pm 150\Gamma$, and we use ($\Gamma_3^N/\Gamma,\Gamma\tau$)$=$$(5, 0.25)$, where $\Gamma$ is the intrinsic decay rate of the idler transition.}\label{fig2}
\end{figure}

In Fig. \ref{fig2}(a), we first explore the effect of the cavity modulation in multiplexing two atomic ensembles, where either the first signal and the second idler photons or both the idler photons go through the optical cavity. The cavity linewidths can be quantified as $\Gamma_c^{s1}$, $\Gamma_c^{i2}$ , and $\Gamma_c^{i1}$, respectively. These two effective spectral functions can be expressed respectively as
\bea
f_{A}&&=f(\omega_s,\omega_i)\left[C_t(\Delta\omega_s,\Gamma_c^{s1})+ C_t(\Delta\omega_i,\Gamma_c^{i2})\right],\\
f_{B}&&=f(\omega_s,\omega_i)\left[C_t(\Delta\omega_i,\Gamma_c^{i1})+ C_t(\Delta\omega_i,\Gamma_c^{i2})\right] .
\eea
In both cases, $S$ is enhanced as cavity modulates either signal or idler photons or both of them. We find a saddle-like distribution in $S$ for $f_A$, while $S$ of $f_B$ always increases compared to the cavity-free case. This shows different behaviors for cavity modulation in the first signal or idler photons when another finite cavity modulation is applied in the second biphoton source. The entanglement entropy of $f_B$ approaches the cavity-free case when $\Gamma_c^{i1}=\Gamma_c^{i2}$ owing to the common phase modulation. This can be regarded as a global phase to the spectral function and thus does not modify the entanglement property. By contrast, the saddle-like behavior for the case of $f_A$ provides a controllable way to suppress $S$ from increasing definitely, which can be seen in Fig. \ref{fig2}(b). 

\begin{figure}[b]
\centering
\includegraphics[width=8.5cm,height=4.5cm]{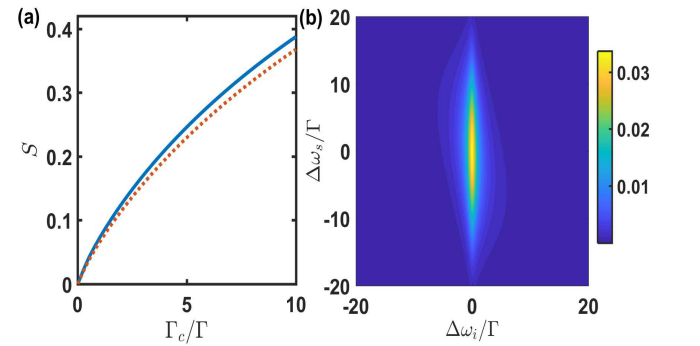}
\caption{Entanglement reduction via cavity and phase modulations. (a) The entanglement entropy $S$ of either one idler (solid-blue) or signal (dotted-orange) going through the cavity ($\Gamma_c^{i2}=\Gamma_c$) with a relative phase $\phi=\pi$, which decreases as $\Gamma_c$ is reduced. The corresponding spectral function of $f_C$ is demonstrated in (b) at $\Gamma_c/\Gamma=1$. The rest of the parameters are the same as in Fig. \ref{fig2}.}\label{fig3}
\end{figure}

In Fig. \ref{fig2}(c), we further explore the joint effect of cavity and phase modulations. The effective spectral function can be written as
\bea
f_{C}=f(\omega_s,\omega_i)\left[e^{i\phi}+ C_t(\Delta\omega_{i(s)},\Gamma_c^{i2(s2)})\right],\label{fc}
\eea
where the second biphoton state can be modulated via cavity on either idler or signal photons. For the same cavity linewidth $\Gamma_c^{i2}=\Gamma_c^{s2}$, we find similar dependence of $\phi$ in $S$ for $f_C$, while as $\phi$ varies around $\pi$, we can identify the minimums of entanglement entropy, which arise from a phase compensation that destructively interferes with the cavity modulation. This can be seen in the spectral distribution in Fig. \ref{fig3}, where the frequency distribution along $\Delta\omega_s+\Delta\omega_i=0$ is suppressed, which is the origin of the frequency entanglement. Specifically, the joint effect makes the distribution in $\Delta\omega_i$ confined within $|\Delta\omega_i|\lesssim \Gamma_c^{i2}$, and the corresponding entanglement entropy $S$ can be reduced and vanishing indefinitely as the cavity linewidth is made negligible, as shown in Fig. \ref{fig3}(a). We note that, however, the cavity linewidth always has a lower bound from thermal fluctuations, which would limit the capability of spectral manipulations in reducing $S$. 

\begin{figure}[t]
\centering
\includegraphics[width=8.5cm,height=4.5cm]{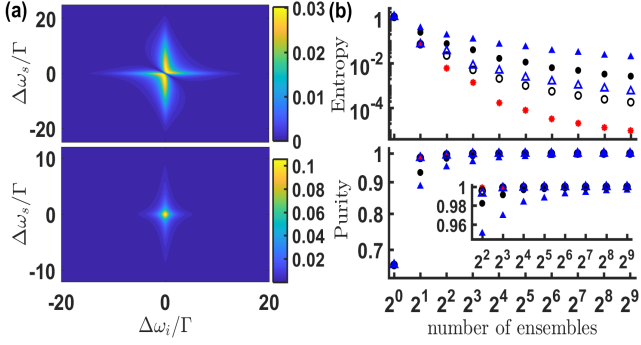}
\caption{Multiplexing multiple atomic ensembles. (a) The upper plot shows the spectral function distribution of $f_D$ for the case of both signal and idler photons from the same AE going through respective cavities with $\Gamma_c^{s2(i2)}=\Gamma$ along with a relative $\pi$ phase, while the lower one shows the distribution of $f_E$ using four AEs with a common cavity  linewidth $\Gamma_c=\Gamma$. (b) The entanglement entropy $S$ as a dependence of the number of multiplexing AEs with phase compensation and the associated purity of single photons at $\Gamma_c/\Gamma$ $=$ $1$ ($\ast$), $5$ ($\bullet$), $10$ ($\blacktriangle$). A comparison of the cases with $(\Gamma_c^i,\Gamma_c^s)/\Gamma$ $=$ $(1,5)$ ($\circ$) and $(1,10)$ ($\triangle$) is also plotted for cavity modulations on the idler and signal photons alternatively. The other parameters are the same as in Fig. \ref{fig2}.}\label{fig4}
\end{figure}

As a natural extension to Eq. (\ref{fc}) and exploration for the possibility of reducing $S$, we consider another multiplexing scheme with extra cavity modulation on the signal photon for the case of $\phi=\pi$, where the effective spectral function can be written as
\bea
f_D=f(\omega_s,\omega_i)\left[e^{i\pi}+ C_t(\Delta\omega_{i},\Gamma_c^{i2})C_t(\Delta\omega_{s},\Gamma_c^{s2})\right].
\eea
We plot the above spectral distribution in Fig. \ref{fig4}(a) and find a compressed but rather asymmetric distribution along with a vanishing cut at $\Delta\omega_s+\Delta\omega_i=0$. This multiplexing scheme results in a compressed spectral distribution in the signal photon but with a price to broaden the idler one instead. The $S$ in this scheme is calculated as around $0.8$, which is lower than the cavity-free case but can not be manipulated to an even lower one compared to the case using Eq. ($\ref{fc}$) in Fig. \ref{fig3}(a). 

We can circumvent this detrimental effect from cavity modulations by multiplexing another two atomic ensembles. The idea is to apply the similar destructive interference in Eq. (\ref{fc}) to the signal frequency distribution as well, and we obtain 
\bea
f_E&&=f_{C'} \left[ e^{i\pi}+C_t(\Delta\omega_{s},\Gamma_c)\right], \label{fe1}\\
f_{C'}&&=f(\omega_s,\omega_i)\left[e^{i\pi}+ C_t(\Delta\omega_{i},\Gamma_c)\right],\label{fcprime} 
\eea
which further leads to an effective spectral function involving four atomic ensembles, 
\bea
f_E&&=f(\omega_s,\omega_i)\left[ 1 - C_t(\Delta\omega_{i},\Gamma_c)-C_t(\Delta\omega_{s},\Gamma_c)\right.\nonumber\\
&&\left.+C_t(\Delta\omega_{i},\Gamma_c)C_t(\Delta\omega_{s},\Gamma_c)\right]. \label{fe}
\eea
As shown in Fig. \ref{fig4}(a), a symmetrically compressed profile of Eq. (\ref{fe}) is achieved. The entanglement entropy $S$ is significantly suppressed to $0.006$ and the associated purity gives $0.999$. This shows the capability of multiplexing scheme with cavity and phase modulation in generating the pure single photons, which originates from the more concentrated and symmetric spectral function. We note that the multiplexing scheme utilizes the destructive interference in the frequency modulations with an aid of external cavities, in contrast to a straightforward truncation of spectral weights in the biphoton state, which compromises its generation rate.  

In Fig. \ref{fig4}(b), we further explore the potentiality of the scheme with multiplexing more atomic ensembles. We apply the cavity modulation with the same cavity linewidth $\Gamma_c$ alternatively on the idler and signal photons, as in Eqs. (\ref{fe1}) and (\ref{fcprime}). Although the number of AEs is increased exponentially to have a more reduced entanglement entropy, we note of the improved level of purity up to $0.9995$ when a moderate number of $64$ AEs are used at $\Gamma_c/\Gamma\sim 5$ and more than $0.9999$ when $\Gamma_c/\Gamma\sim 1$. This presents a high-purity single photon source via multiplexing biphotons with cavity modulation and phase compensation. 

\section{Symmetrization of spectral function}

\begin{figure}[b]
\centering
\includegraphics[width=8.5cm,height=4.5cm]{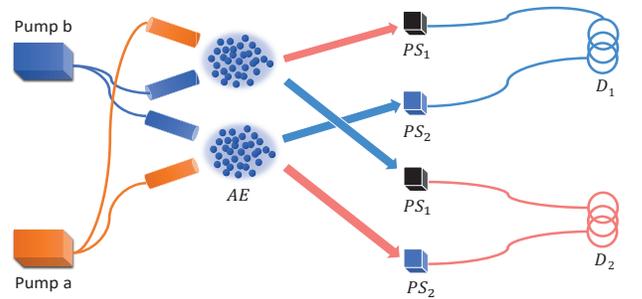}
\caption{A schematic setup for symmetrization of the spectral functions. The directions of pump fields have exchanged for the second AE, and the non-discrimination detectors $D1$ and $D2$ are used along with respective phase shifters (PSs) to achieve the effective symmetrized spectral function.}\label{fig5}
\end{figure}

In the same spirit of reducing $S$ under a more confined and symmetric spectral function in the previous section, here we investigate the continuous entanglement entropy under a symmetrized spectral function with non-discrimination detections. The symmetrization of spectral function can be done easily for spontaneous parametric down conversion process of entangled photons \cite{Branning1999, Law2000}. They are identical in their central frequencies and can be manipulated in their polarizations to achieve the symmetrization. On the other hand in the cascaded configuration of atomic ensembles, the signal and idler photons are in general non-degenerate. Therefore, as shown in Fig. \ref{fig5}, we propose to exchange the excitation fields directions in two AEs, and such that each non-discrimination detector would detect signal and idler photons simultaneously. Under this condition, we obtain effectively a symmetrized spectral function,
\bea
f_s=f(\Delta\omega_1,\Delta\omega_2)+e^{i\phi}f(\Delta\omega_2,\Delta\omega_1), \label{fs}
\eea
where $s$ stands for symmetrization, and the subscripts $1$ or $2$ attached to the frequency distributions are associated to the detectors.   

\begin{figure}[t]
\centering
\includegraphics[width=8.5cm,height=4.5cm]{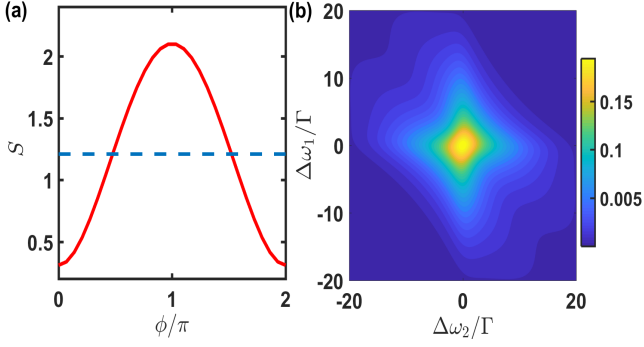}
\caption{(a) The entanglement entropy $S$ from the symmetrization of spectral function (solid-red) on various relative phases, comparing the one without symmetrization (dash-blue). (b) The spectral distributions of symmetric spectral function at ($\Gamma_N^3/\Gamma,\Gamma\tau$) $=$ $(5, 0.25)$.}\label{fig6}
\end{figure}

This symmetrization can be fulfilled either by frequency conversion of the signal field into the one that coincides with idler's central frequency \cite{Radnaev2010, Jen2010}, or use a cascade transition with similar bandwidth for both signal and idler photons, for example the upper transitions of $5$D$_{5/2}$ to $5$P$_{3/2}$ ($776$nm) \cite{Chaneliere2006} or $5$D$_{3/2}$ to $5$P$_{1/2}$ ($762$nm) \cite{Seidler2020} can be accessed in rubidium atoms. In Fig. \ref{fig6}, we show the dependence of $S$ on the modulated phase $\phi$ in Eq. (\ref{fs}), and as expected the entanglement entropy presents a minimum (maximum) when $f_s$ is symmetrized (anti-symmetrized). The calculated $S$ of the minimum is $0.3$ and the associated single photon purity can reach $0.9226$. Although this scheme does not show the advantage over the one with cavity and phase modulations, it can still provide a single photon source with relatively low frequency correlations without much endeavor in applying external optical elements.   

\section{Conclusion}

In conclusion, we have theoretically studied the continuous entanglement entropy and single photon purity of the cascaded biphoton state in a multiplexing scheme. We propose two schemes to reach low entanglement entropy in frequency spaces via multiplexing atomic ensembles with cavity modulation and phase compensation. The associated single photon purity can reach the level of $0.999$, which suggests that an almost indistinguishable single photon can be generated by removing the mutual frequency correlations in the cascaded biphoton source. We show that it can be done by a destructive interference between the cavity and phase modulations, which leads to a symmetric and confined spectral function without sacrificing the generation rates. The other approach employs a symmetric pumping of the laser fields in two atomic ensembles, which also provides a moderate entropy reduction when non-discrimination detectors are applied. Our results provide an alternative resource to access near-unity photon purity, which is essential in photonic quantum technologies \cite{Feng2017}. In addition, a near-unity purity of single photon is also indispensable, for example, in Gaussian Boson sampling experiment \cite{Zhong2021} to demonstrate the capability of photonic quantum computer to surpass various classical simulation schemes. 

\section*{Acknowledgments}

This work is supported by the Ministry of Science and Technology (MOST), Taiwan, under the Grant No. MOST-109-2112-M-001-035-MY3. H.H.J. is also grateful for support from TG 1.2 and TG 3.2 of NCTS.
\appendix
\section{Biphoton state from a cold atomic ensemble}

Here we show the detail in deriving the spectral function of the biphoton state from a cold atomic ensemble, which has been investigated in an atomic ensemble with a diamond-type configuration \cite{Jen2016a, Jen2017_cascaded} as shown in Fig. \ref{fig1}. The biphoton state is generated from a cold atomic ensemble driven by two weak laser fields with Rabi frequencies $\Omega_a$ and $\Omega_b$ under four wave mixing condition. The interaction Hamiltonian reads 
	\bea
	\label{InteractionH}
	V_I &&= - \sum_{m=1,2} \Delta_m \sum_{\mu=1}^{N} \ket{m}_\mu \bra{m} - \sum_{m=a,b} \left(\frac{\Omega_m}{2} \hat{P}_m^\dagger + h.c.\right) \nonumber\\
		&&-i \sum_{m=s,i} \left[ \sum_{\mathbf{k}_m, \lambda_m} g_m \hat{a}_{\mathbf{k}_m, \lambda_m} \hat{Q}_{m}^{\dagger} e^{-i\Delta \omega_m t} - h.c. \right],
	\eea
	where $\hbar=1$, the polarization is $\lambda_{m}$, the wave vectors are $\mathbf{k}_{m}$, and the detunings are defined as $\Delta_1 = \omega_{a} - \omega_{1}$, $\Delta_2 = \omega_{a} + \omega_{b} - \omega_{2}$ with atomic level energies $\omega_{m=1,2,3}$ and central frequencies $\omega_{a,b,s,i}$. We have incorporated $\epsilon_{\mathbf{k}_{m}, \lambda{m}} \cdot \hat{d}_{m}^{*}$ into the biphoton coupling constants $g_m$ with the polarization direction $\epsilon_{\mathbf{k}_m, \lambda_{m}}$ of the quantized bosonic fields $\hat{a}_{\mathbf{k}_m, \lambda_{m}}$ and the unit direction of dipole operators $\hat{d}_{m}$. Dipole operators are defined as $\hat{P}_a^\dagger \equiv \sum_\mu \ket{1}_\mu \bra{0} e^{i \textbf{k}_a \cdot \textbf{r}_\mu }$, $\hat{P}_b^\dagger \equiv \sum_\mu \ket{2}_\mu \bra{1} e^{i \textbf{k}_b \cdot \textbf{r}_\mu }$, $\hat{Q}_s^\dagger \equiv \sum_\mu \ket{2}_\mu \bra{3} e^{i \textbf{k}_s \cdot \textbf{r}_\mu }$, $\hat{Q}_i^\dagger \equiv \sum_\mu \ket{3}_\mu \bra{0} e^{i \textbf{k}_i \cdot \textbf{r}_\mu }$.
	
Under weak field approximation with large detunings, that is $\sqrt{N} |\Omega_{a}|$$/$$\Delta_{1}$$\ll$$1$ \cite{Jen2012-2}, we assume single excitation regime, and the state function can be written as
	\bea
	|\Psi(t)\rangle &&= \mathcal{E}(t) |0, {\rm vac}\rangle + \sum_{\mu=1}^{N} A_{\mu}(t) |1_{\mu}, {\rm vac}\rangle\nonumber\\
	    &&+ \sum_{\mu=2}^{N} B_{\mu}(t) |2_{\mu}, {\rm vac}\rangle + \sum_{\mu=1}^{N} \sum_{s} C_{s}^{\mu}(t) |3_{\mu}, 1_{\mathbf{k}_{s}, \lambda_{s}}\rangle \nonumber\\
	    &&+ \sum_{s,i}D_{s,i}(t) |0, 1_{\mathbf{k}_{s}, \lambda_{s}}, 1_{\mathbf{k}_{i}, \lambda_{i}}\rangle,
	\eea
	where the collective single excitation states and the vacuum photon state are $\ket{m_{\mu}} \equiv \ket{m_{\mu}} \ket{0}_{\nu \neq \mu}^{\otimes N -1}$ and $\ket{{\rm vac}}$ respectively.
	
	Using Schr\"{o}dinger equation $i\frac{\partial}{\partial t}|\Psi(t)\rangle$$=$$V_I|\Psi(t)\rangle$, we have the following equations of motion,
	\begin{align}
		i \dot{\mathcal{E}} &= - \frac{\Omega^\ast_a}{2} \sum_\mu e^{-i \textbf{k}_a \cdot \textbf{r}_\mu} A_\mu, \\
		i \dot{A}_\mu &= - \frac{\Omega_a}{2} e^{i \textbf{k}_a \cdot \textbf{r}_\mu} \mathcal{E} - \frac{\Omega^\ast_b}{2} e^{-i \textbf{k}_b \cdot \textbf{r}_\mu} B_\mu - \Delta_1 A_\mu, \\
		i \dot{B}_\mu &= - \frac{\Omega_b}{2} e^{-i \textbf{k}_b \cdot \textbf{r}_\mu} A_\mu - \Delta_2 B_\mu \nonumber\\
		& - i \sum_{k_s, \lambda_s} g_s e^{i \textbf{k}_s \cdot \textbf{r}_\mu} e^{-i (\omega_s - \omega_{23} - \Delta_2) t} C^\mu_s, \\
		\dot{C^\mu_{s,i}} &= i g^\ast_s e^{-i \textbf{k}_s \cdot \textbf{r}_\mu} e^{i (\omega_s - \omega_{23} - \Delta_2) t} B_\mu \nonumber\\
		&- i \sum_{k_i, \lambda_i} g_i e^{i \textbf{k}_i \cdot \textbf{r}_\mu} e^{-i (\omega_i - \omega_3) t} D_{s,i}, \\
		i \dot{D_{s,i}} &= i g^\ast_i \sum_\mu e^{-i \textbf{k}_i \cdot \textbf{r}_\mu} e^{i (\omega_i - \omega_3) t} C^\mu_s.
	\end{align}
	Under the weak field approximation, $\mathcal{E} \approx 1$ for the zeroth order perturbation, and the steady-state solutions of $\dot{A_{\mu}}$$=$$\dot{B_{\mu}}$$=$$0$ give $A_\mu(t) \approx - \Omega_a(t)e^{i\textbf{k}_a \cdot \textbf{r}_\mu} / (2\Delta_1)$, $B_\mu(t) \approx \Omega_a(t)\Omega_b(t) e^{i(\textbf{k}_a + \textbf{k}_b) \cdot \textbf{r}_\mu} / (4\Delta_1 \Delta_2)$, in the leading order of perturbations respectively.
	
	We further consider a symmetrical single excitation state,\\ $(\sqrt{N})^{-1} \sum^N_{\mu=1} e^{i(\textbf{k}_a + \textbf{k}_b - \textbf{k}_s) \cdot \textbf{r}_\mu} \ket{3}_\mu \ket{0}^{\otimes N - 1}$, which leads to the biphoton state $\ket{1_{\textbf{k}_s}, 1_{\textbf{k}_i}}$ generation in a large $N$ limit,
	\bea
	D_{s,i} (t) &&= g^\ast_i g^\ast_s \sum^N_{\mu=1} e^{i\Delta \textbf{k} \cdot \textbf{r}_\mu} \int_{-\infty}^{t} \int_{-\infty}^{t'} dt'' dt' \bigg[e^{i\Delta \omega_i t'}\nonumber\\
	&&\times e^{i\Delta \omega_s t''}\frac{\Omega_a(t'') \Omega_b(t'')}{4\Delta_1 \Delta_2} e^{(-\Gamma^N_3 / 2 + i\delta \omega_i)(t' - t'')} \bigg].\label{integral}\nonumber\\
	\eea
We next assume Gaussian pulse excitations $\Omega_{a,b}(t)$$=$$\tilde{\Omega}_{a,b} e^{-t^2 / \tau^2} / (\sqrt{\pi} \tau)$ with the pulse duration $\tau$ and quantified the superradiant decay rate of the idler photon as $\Gamma_3^N$$=$$(N \bar{\mu} + 1)\Gamma$ \cite{Dicke1954, Lehmberg1970, Gross1982, Jen2012, Jen2015} with an intrinsic decay rate $\Gamma$ and geometrical constant $\bar{\mu}$ \cite{Rehler1971}. This $\Gamma^N_3$ depends on the atomic density and its geometry, which therefore can be tailored and controlled in system preparations. With four-wave mixing condition $\mathbf{k_a + k_b = k_s + k_i}$, $\sum^N_{\mu=1} e^{i\Delta \textbf{k} \cdot \textbf{r}_\mu}$ becomes maximal and corresponds to the phase-matched and highly correlated biphoton state. Finally, after integrating out Eq. (\ref{integral}), we obtain
	\begin {align}
	D_{si} (\Delta \omega_s, \Delta \omega_i) &= \frac{\tilde{\Omega_a} \tilde{\Omega_b} g_i^* g_s^*}{4 \Delta_1 \Delta_2} \frac{\sum_\mu e^{i\Delta \textbf{k} \cdot \textbf{r}_\mu}}{\sqrt{2 \pi} \tau} f(\omega_s, \omega_i),\\
	f(\omega_s, \omega_i) &\equiv \frac{e^{-(\Delta \omega_s + \Delta \omega_i)^2 \tau^2 / 8}}{\frac{\Gamma_3^N}{2} - i \Delta \omega_i}, \label{fc_co}
	\end{align}
where $\Delta \omega_s \equiv \omega_s - (\omega_2 - \omega_3 +\Delta_2)$ and $\Delta \omega_i \equiv \omega_i - \omega_3$. The joint Gaussian distribution in $f(\omega_s, \omega_i)$ maximizes when $\Delta \omega_s + \Delta \omega_i = 0$, which shows the energy conservation of laser excitations and biphoton generations, that is $\omega_s + \omega_i = \omega_a + \omega_b$. This also features an anti-correlation between $\Delta \omega_s$ and $\Delta \omega_i$ and the origin of frequency entanglement entropy.

\end{document}